\documentstyle[12pt]{article}
\date{~}
\title{%
Supermassive Black Holes \\ and the Evolution of Galaxies
\footnote{Almost identical to the text in Nature {\bf 395,}
 A14, Oct 1, 1998}
}

\author{D. Richstone, E. A. Ajhar, R. Bender, G. Bower, A. Dressler, \\
S. M. Faber, A. V. Filippenko, K. Gebhardt, R. Green, L. C. Ho, \\ 
J. Kormendy, T. Lauer, J. Magorrian, S. Tremaine}

\begin{document}

 \def\today{\number\day\enspace
      \ifcase\month\or January\or February\or March\or April\or May\or
      June\or July\or August\or September\or October\or
      November\or December\fi \enspace\number\year}
 \def\clock{\count0=\time \divide\count0 by 60
     \count1=\count0 \multiply\count1 by -60 \advance\count1 by \time
     \number\count0:\ifnum\count1<10{0\number\count1}\else\number\count1\fi}
 \def\datestamp{{\hss \today \quad\quad \clock}}
 \def\newline{\hfil\break}
\def\new{{\bf NEW: \rm }}
\def\gtsim{{_>\atop{^\sim}}}
\def\ltsim{{_<\atop{^\sim}}}
\def\et{{\it et~al. }}
\def\eg{{\it eg: }}
\def\avL{\langle \ell \rangle}
\def\avldet{\langle l_{det} \rangle}
 \def\solar{\ifmmode_{\mathord\odot}\else$_{\mathord\odot}$\fi}  
 \def\bul{\ifmmode\bullet\else$\bullet$\fi}  
 \def\avg#1{\langle #1 \rangle}
\def\xvec{{\bf x}}
\def\vvec{{\bf v}}
\def\avec{{\bf a}}
\def\mass{{\cal M}} 

\def\grad{{\nabla}}
\def\del{{\nabla}}

 \def\deg{\ifmmode^\circ\else$^\circ$\fi}
\def\arcsec{''\hskip-3pt .}
\def\arcmin{'\hskip-2pt .}
 \def\kms{\ifmmode \hbox{ \rm km s}^{-1} \else{ km s$^{-1} $}\fi} 

\def\sec{\ifmmode \hbox{\rm sec}\else{sec}\fi} 
\def\yr{\ifmmode \hbox{\rm yr}\else{yr}\fi} 
\def\myr{\ifmmode \hbox{\rm Myr}\else{Myr}\fi} 
\def\gyr{\ifmmode \hbox{\rm Gyr}\else{Gyr}\fi} 

\def\hz{\ifmmode \hbox{\rm Hz}\else{hz}\fi} 
 
\def\kpc{\ifmmode \hbox{\rm  kpc}\else{kpc}\fi} 
 \def\pc{\ifmmode \hbox{\rm  pc} \else{pc}\fi} 
\def\mpc{\ifmmode \hbox{\rm Mpc} \else{Mpc}\fi} 

\def\erg{\ifmmode \hbox{\rm erg} \else{erg}\fi} 

 \def\yrs{\ifmmode \hbox{\rm yrs}\else{yrs}\fi} 
 \def\Msun{M_\odot}
 \def\mbh{M_\bullet}
 \def\lbulge{L_{bulge}}
 \def\mbulge{M_{bulge}}
 \def\Lsun{L_\odot}
 \def\rsun{r_\odot}
 \def\angstr{\ifmmode{\rm \AA}\else\AA\fi}

\def\ho{\ifmmode H_0\else$H_0$\fi}
\def\omo{\ifmmode\Omega_0\else$\Omega_0$\fi}
\def\to{\ifmmode T_0\else$T_0$\fi}
\def\h-1{h^{-1}}
\def\rhobar{\langle\rho\rangle}
\def\dd#1#2{{d #1 \over d #2 }}  
\def\d2#1#2{{d^2#1 \over d#2^2}}
\def\rdot{\dot r}
\def\rddot{\ddot r}


\def\apj{ApJ}
\def\apjl{ApJ Letters}
\def\Apj{ApJ}
\def\aj{AJ}
\def\aa{Astron \& Astroph}
\def\annrev{Ann Rev Astron and Astroph}
\def\mn{MNRAS}
\def\pasp{Pub.~A~S~P}

\def \lap
 {\mathrel{\hbox{\raise0.3ex\hbox{$<$}\kern-0.75em\lower0.8ex\hbox{$\sim$}}}}

\def\new{{\bf NEW: \rm }}

\input psfig.tex

\maketitle

\renewcommand\baselinestretch{1.0}
\small\normalsize

\bigskip
\bigskip

{\bf Black holes, an extreme consequence of the mathematics of General
Relativity, have long been suspected of being the prime movers of
quasars, which emit more energy than any other objects in the
Universe.  Recent evidence indicates that supermassive black holes,
which are probably quasar remnants, reside at the centers of most
galaxies.  As our knowledge of the demographics of these relics of a
violent earlier Universe improve, we see tantalizing clues that they
participated intimately in the formation of galaxies and have strongly
influenced their present--day structure.}

\bigskip
\bigskip

Black holes are a prediction of Einstein's theory of
gravity, foreshadowed by the work of Michell and later Laplace in the
late 18th century.  K.  Schwarzschild discovered the simplest kind  
of black hole in the first solution of Einstein's equations of General
Relativity, and Oppenheimer was among the first to consider the
possibility that black holes might actually form in nature.  The
subject gained life in the 1960s and 70s, when supermassive black
holes were implicated as the powerhouses for quasars and stellar--mass
black holes were touted as the engines for many galactic X--ray sources.  In
the last decade, we have progressed from seeking supermassive black
holes in only the most energetic astrophysical contexts, to suspecting
that they may be routinely present at the centers of galaxies
\cite{begelrees, kr, kafatos}. 

The defining property of a black hole is its event horizon.
Since the horizon itself is invisible, we must often settle for
evidence of mass without light.  All dynamical techniques for finding
supermassive black holes at the centers of galaxies rely on a
determination of mass enclosed within a radius $r$ from the velocity
$v$ of test particles; in Newtonian physics, this mass is $M_r =
\alpha \, v^2 r /G$.  Determining $\alpha$ requires a detailed
dynamical analysis, but it is often of order $1$.  In
cases where there is extra mass above that associated with starlight,
we refer to the object as a ``massive dark object'' (or MDO).  In most
of the cases discussed in this paper, it is likely that the 
MDO is a supermassive black hole (MBH), but in only a few cases have
plausible alternatives to a black hole been ruled out.  These are 
important as they establish the reality of MBH and justify 
the  interpretation of less compelling objects as MBHs.

\section{Black holes as the energy sources of quasars}

Black holes are thought to exist in two mass ranges. Small ones of
$\sim 10 \Msun$ are the evolutionary end points of some massive stars.
This paper discusses the much more massive ones that might power
quasars and their weaker kin, active galactic nuclei (AGN).  Quasars
produce luminosities of $L \sim 10^{46}$ ergs/sec ($\sim
10^{12}\Lsun$).  Where they power double--lobed radio sources, the
minimum energy stored in the lobes is $E \sim 10^{60 - 64}$ ergs. The
mass equivalent of this energy is $ M = E/c^2 \sim 10^{6 - 10} \Msun
$, and the horizon scale associated with that mass is $ R_S = G M/c^2
\sim 10^{11 - 15}$ cm.  Although most quasars do not vary much at
visual wavelengths, a few objects change their luminosity in minutes
at high energies \cite{barr, edelson}.  Since an object cannot
causally vary faster than the light--travel time across it, such
objects must be smaller than $R \sim c \tau \sim 10^{13}$ cm.
Although relativistic corrections can alter this limit somewhat in
either direction via Doppler boosting or gravitational redshift, there
is no escaping the conclusion that many quasars are prodigiously 
luminous yet tiny, outshining a galaxy in a volume smaller than the 
solar system.

The small size together with the enormous energy output of quasars
mandates black hole accretion as the energy source.  
Most investigators believe that quasars and AGNs are MBHs accreting
mass from their environment, nearly always at the center of a galaxy
\cite{lbnature, blandagn, abrambh}.  Black holes of mass $ >10^7 \Msun$ must
normally lie at the center because dynamical friction drags them to
the bottom of the potential well.  This location is now clearly
established for low redshift ($z \ltsim 0.3$) quasars \cite{bkss}.
The connection between MBHs and quasars was first made by Zeldovich
\cite{zeld} and Salpeter \cite{salp}. Lynden-Bell \cite{lbscripta} sharpened
the argument by computing the ratio of gravitational energy to nuclear
energy
\begin{equation}
{ E_g \over E_n}\sim {\epsilon _g \, GM^2/R \over \epsilon _n \, M c^2}
      \sim  
      \left({ \epsilon _g \over \epsilon _n}\right) 
      \left({R_S \over R}\right)
       \sim 100 \epsilon_g, 
\end{equation}
where $R_S$ is the Schwarzschild radius of a black hole of mass $M$,
$R$ is the size of the quasar, and $\epsilon _g$ and $\epsilon _n$ are
gravitational and nuclear energy conversion efficiencies; the last
equality follows from the typical astrophysical thermonuclear
efficiency of $ \sim 1\% $ and the size scale from variability noted
above.  

Because quasars were populous in the youthful Universe, but 
have mostly died out, the Universe should be populated with 
relic black holes whose 
average mass density $\rho_u$ matches or exceeds the mass--equivalent of
the energy density $u$ emitted by them \cite{chokshi}. 
The integrated comoving energy density in quasar light (as emitted)
is 
\begin{equation}
u 
= \int_0^\infty \int_0^\infty \, L \Phi(L|z) dL \, {dt \over dz} \, dz
= 1.3 \times 10^{-15} {\rm erg \; cm^{-3}},  
\end{equation}
where $\Phi$ is the comoving density of quasars of luminosity $L$, 
and $t$ is cosmic time.  The corresponding present-day mass density
for a radiative efficiency $\epsilon$ is 
$\rho_u = {u / [\epsilon c^2]} = 
           2 \times 10^5  
            \left({ 0.1 \over \epsilon} \right) \Msun \, \mpc^{-3}.  
$
This density can be compared to the luminous density in galaxies,
$j = 1.1 \times 10^8 \, \Lsun \, \mpc^{-3}$ 
\cite{loveday}, to obtain the ratio of the mass  
in relic MBHs to the light of galaxies:
\begin{equation}
  \Upsilon = 
     {\rho_u \over j} = 
      1.8 \times 10^{-3} \; \left( {0.1 \over \epsilon} \right)
         \left( {\Msun \over \Lsun} \right) . 
\end{equation}

\section{Dynamical evidence for massive black holes}

\subsection{First steps}

The first dynamical evidence for black holes in galactic centers was
the 1978 measurement \cite{sargentyoung} of a rising
central velocity dispersion, reaching $\sim 400 \kms$, in the giant
elliptical galaxy M87.  This object is a prime site to prospect for 
an MBH by virtue of its AGN features --- nonthermal radio emission, broad
nuclear emission lines, and a ``jet'' of collimated relativistic
particles being ejected from the nucleus.  Isotropic models of the
stellar kinematics, when combined with photometry, implied an MDO of
$5 \times 10^9 \Msun$.  The result was criticized  
because the data were also matched by a model with
{\it radially anisotropic} stellar orbits and no black hole.  Thus,
the importance of understanding the stellar orbital structure of
the centers of galaxies was obvious at the very beginning, and this
subject has developed in parallel with the search for MBHs.

More convincing evidence was found in the 1980s for MDOs in M31 and M32 (the
Andromeda galaxy and its satellite), which are nearby and hence 
observable at high spatial resolution \cite{drm31, kormm31}. 
Rapid rotation near their centers reduces the danger of confusing a
central mass with radial orbits, and Schwarzschild's method of loading
orbits in a galaxy potential \cite{schmeth, maxent1} was used to
eliminate unphysical models. Modern methods use Schwarzschild's method
to fit the entire line--of--sight velocity distribution for
axisymmetric models.

The stellar velocity work on M87 was largely vindicated two decades
later by the {\it Hubble Space Telescope (HST)}, which revealed a
small gas disk at the center \cite{harmsm87,fordm87}.  The gas is
plausibly in circular motion, so the MDO mass estimate is
straightforward.  These and later data provide strong evidence for an
MDO of $3 \times 10^9 \Msun$ \cite{harmsm87,maccm87}, a value
similar to but slightly smaller than the one derived in
\cite{youngsargent}.

\subsection{Two remarkable examples}

The work described above revealed strong examples of MDOs but
no iron--clad evidence for MBHs.  This gap has now been
partly closed in two remarkable objects.  The mild AGN NGC 4258 was shown
\cite{miyoshi} to possess a tiny annular gaseous disk near the
nucleus, populated by water masers whose Doppler velocities can be
observed with exquisite precision.  The rotation curve is Keplerian
to high accuracy over the annulus width ($0.13-0.26 \, \pc$).  The very
small velocity residuals of $\ltsim 1\%$ inspire confidence in the
derived mass of $3.6 \times 10^7 \Msun$ \cite{maoz}.  The extraordinarily high
implied density of $> 10^9 \ \Msun \pc^{-3} \; (10^{12}$ if one takes
the limits on departures from Keplerian motion as a constraint on the
concentration of the mass) permits the use of astrophysical arguments
to rule out most other explanations for the dark mass (see
below). This is a firm link from MDOs to MBHs.

The center of our Galaxy holds the second confirmed MBH.
Near--infrared observations detect {\it proper motions} of stars in
orbit about the galactic center \cite{genzel, ghez} and indicate a
rising stellar velocity dispersion down to distances of 0.01 pc.  For
the first time stars are being observed to orbit an MDO, year by year,
with impressive accuracy that will steadily improve.  The density of
$>10^{12} \, \Msun \ \mpc^{-3}$ within the resolved region is again
extraordinarily high, ruling out most alternatives to an MBH.

In the Galaxy and NGC 4258, the MDOs are almost surely MBHs
\cite{maoz} rather than clusters of smaller masses.  A cluster of
radius $r$ and total mass $M$ of $N$ self--gravitating point masses
will collapse or evaporate on a timescale of a few hundred two--body
relaxation times, $t_r = 0.14 N \, (r^3/GM)^{1/2} \, [\ln
(0.4N)]^{-1}$ (some important caveats were noted in \cite{goodlee}).
At a chosen nonrelativistic density, the lifetime of the cluster can
be made longer than the age of the Universe by making the point masses
sufficiently light (and therefore numerous).  In NGC 4258 and the
Galaxy, this constraint requires constituents with masses $\ltsim 0.1
\Msun$.  Brown dwarfs and white dwarfs with this mass have large radii
and would rapidly collide and merge.  Thus the remaining candidate
components for cluster models are low--mass black holes or
noninteracting elementary particles.  However, there is no known way
to make $0.1 \Msun$ black holes, and, since non--interacting elementary
particles do not radiate energy efficiently they do not settle into a
compact configuration.  The cases for black holes in both NGC 4258 and
the Galaxy therefore seem very strong.  The argument for all of the
other MDOs is weaker, but, by analogy and by virtue of the
overwhelming circumstantial evidence for MBHs in AGNs they are now the
preferred explanation.

\subsection{Potential new tools}

The techniques above are difficult to apply to galaxies containing
AGNs.  The bright nucleus renders the stellar absorption lines close
to the center nearly invisible, and in many cases the nebular emission
lines are influenced by nongravitational forces.  Nevertheless, in
AGNs that emit broad emission lines originating from gas near the
central engine ($\Delta v \gtsim 10,000 \kms$), 
one can attempt to estimate a mass from the average
velocity of the gas and the radius of the emitting region.  The
velocity comes from the widths of the lines, but the radius is harder
to measure.  It can be estimated either from photoionization models of
the gas or by ``reverberation mapping.''  In the latter method, the
radius is inferred from the time interval (due to light travel)
between fluctuations in the continuum radiation and the changes these
induce in the emission lines \cite{netzpeter}.

Masses obtained from reverberation mapping for nearby Seyfert nuclei
range from $10^7$ to $10^8 \Msun$ \cite{ho98}, roughly
consistent with (but somewhat smaller than) the dynamical results of
the previous section.  However, the identifications of line width with
orbital velocity and time delay with radius are problematic given the
absence of any correlation between line width and radius of the form
$v \propto r^{-1/2}$ within the same object.  Evidently some essential
component of the model is still missing.  A proper understanding of
this technique would give us a powerful tool for more luminous and
distant objects \cite{kaspi}.

Recent advances in X--ray astronomy have furnished dramatic new evidence for
MBHs in AGNs.  It had been known for some time that
the X-ray spectra of many AGNs show an iron K$\alpha$ emission line at
a rest energy of 6.4 keV, thought to arise from X--ray
fluorescence of cold, neutral material in an accretion disk. 
 Until recently, the available  spectral resolution 
was insufficient to test the predicted line profile, 
but the {\it Advanced Satellite for Cosmology and 
Astrophysics}  provided the much--awaited
breakthrough in the Seyfert 1 galaxy
MCG--6-30-15 \cite{tanaka}. The K$\alpha$ line exhibits
relativistic Doppler motions of nearly 100,000 km s$^{-1}$, as well as
an asymmetric red wing consistent with gravitational redshift.  The
best--fitting disk has an inner radius of only a few Schwarzschild
radii.  The Fe K$\alpha$ line profile has now been seen in
many objects \cite{nandra}, and data of better quality may eventually
even allow measurement of the spin of the black hole
\cite{reynoldsbeg,bromley}.

\section{The demographics of supermassive central black holes}

\begin{figure}
\vskip 0in
\hskip 0.3in\vbox{\psfig{figure=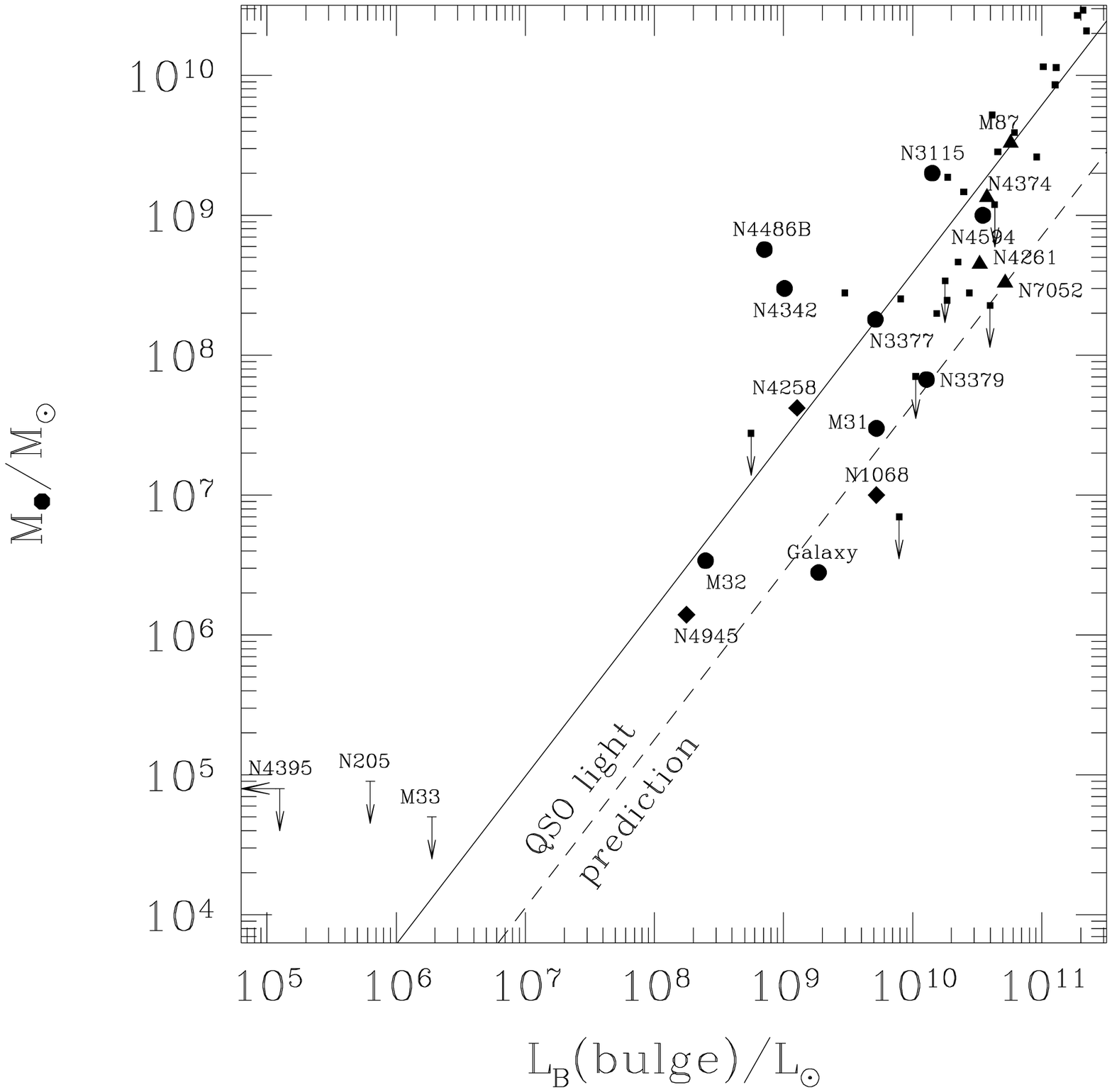,height=5.in,angle=-0.}}
\vskip -0.25in

Figure 1. ---

Mass estimates of the candidate MBHs in galaxies with dynamical
information plotted against the bulge luminosity of their host galaxy.
The labeled points are the results of painstaking observation and
detailed modelling. 
The symbols indicate the how $\mbh$ was derived: 
kinematics of gas --- triangles; dynamics of stars --- filled circles;
masers --- diamonds; or two-integral modelling using ground-based
stellar kinematics --- small squares.  Arrows indicate upper limits on
$\mbh$.  The solid line is a model with $\mbh = 0.005M_{bulge}$ 
and . $ M_{bulge} = 5 (L_{bulge}/10^9 \Lsun)^{1.2}$.  
The distribution of ~$\mbh$ is roughly Gaussian in
$\log (\mbh/M_{\rm bulge})$ with mean $-2.27$ ($\mbh/M_{bulge} =
0.005$) and standard deviation 0.5. 
The dashed line is the quasar light prediction of eqn 3 apportioned
according to the bulge mass: $ \mbh = 2 \times 10^7 ( L_{bulge}/5
\times 10^9 \Lsun)^{1.2}$.

The small offset from the observed
black-hole/bulge-mass relation indicates that the present integrated
density in MBHs is broadly consistent with the integrated
luminosity produced by AGNs over the life of the Universe.  This
offset may reflect a radiative efficiency of average quasar accretion
less than 0.10.

\end{figure}

We now turn to the questions of the number of MBHs in the 
Universe.  Figure 1 illustrates the relationship  
between black hole mass and host spheroid luminosity from the data in
Table 1 (labeled points).  The labeled MDO masses seem
to correlate with spheroid luminosity (solid line); the upper limits are
also consistent with this relation \cite{kr, k93madrid}.  
However, the number of points is small, and,  further progress
requires more objects.  Few objects at
the present time have been studied with the detailed spatial
resolution and/or modeling of the labeled points.

\begin{minipage}{5.in}
\centerline{DYNAMICALLY IDENTIFIED MDOs}
\hskip 150pt{
\begin{tabular}{llcccrl}
\noalign{\smallskip}
\hline
\hline
\noalign{\smallskip}
~Galaxy               &
Name                 &
Type                 &
Distance \footnote{All quantities in this paper are computed for a
Friedman-Robertson-Walker Universe with $\Omega = 1 $ and $\ho = 80
\kms \mpc ^{-1}$. Distances to nearby MBHs come from many
sources, but are always rescaled to this Hubble constant. }    &
M$_{\rm B}$          &
M$_\bullet$ ~~~         &
Reference            \\
         &         &             & Mpc     & Bulge   &   M$_{\rm sun}$~~~   &           \\
\noalign{\smallskip}
\hline
\noalign{\smallskip}
\multicolumn{7}{c}{Stellar Dynamics} \\
~~~------&  Galaxy &  Sbc        &  0.0085 & --17.70 &  $2.8\times10^6$  &  see \cite{genzel, ghez} \\
NGC0221  &  M32    &  E2         &  0.7    & --15.51 & $3.4\times10^6$&  \cite{drm31,tonry84, benderm32, vdmm32}\\
NGC0224  &  M31    &  Sb         &  0.7    & --18.82 &  $3.0\times10^7$  &  \cite{drm31,kormm31}  \\
NGC3115  &  ---    &  S0         &  8.4    & --19.90 &  $2.0\times10^9$  &  \cite{krn3115,knuk3115} \\
NGC3377  &  ---    &  E5         &  9.9    & --18.80 &  $1.8\times10^8$  &  \cite{korm3377} \\
NGC3379  &  M105   &  E1         &  9.9    & --19.79 &  $6.7\times10^7$  &  \cite{geb3379} \\
NGC4342  &  IC3256 &  S0         & 15.3    & --17.04 &  $3.0\times10^8$  &  \cite{vdbosch97}   \\
NGC4486B &  ---    &  E1         & 15.3    & --16.65 &  $5.7\times10^8$  &  \cite{knuk4594} \\
NGC4594  &  M104   &  Sa         &  9.2    & --20.88 &  $1.0\times10^9$  &  \cite{knuk4594} \cite{kormn4594}\\
\hline
\multicolumn{7}{c}{Gas Dynamics} \\
NGC4374  &  M84    &  E1         & 15.3    & --20.96 & $1.4\times10^9$   &  \cite{bowerm84} \\
NGC4486  &  M87    &  E0         & 15.3    & --21.42 & $3.3\times10^9$   &  \cite{harmsm87,fordm87} \\
NGC4261  &  ---    &  E2         & 27.4    & --20.82 & $4.5\times10^8$   &  \cite{ferrn4261} \\
NGC7052  &  ---    &  E4         & 58.7    & --21.31 &  $3.3\times10^8$  &  \cite{vdmn7052}      \\
\hline
\multicolumn{7}{c}{Maser Dynamics} \\
NGC1068  &  M77    &  Sb         & 15.     & --18.82 & $1.0\times10^7$   &  \cite{grnhill1068} \\
NGC4258  &  M106   &  Sbc        &  7.5    & --17.28 & $4.2\times10^7$   &  \cite{miyoshi} \\
NGC4945  &  ---    &  Scd        &  3.7    & --15.14 &  $1.4\times10^6$  &  \cite{grnhill4945} \\
\hline
\multicolumn{7}{c}{Upper Limits} \\
NGC0205  &  ---    &  Spheroidal &  0.72   & --9.02  & $<9.\times10^4$   &  \cite{jones} \\
NGC0598  &  M33    &  Scd        &  0.795  & --10.21 & $<5.\times10^4$   &  \cite{kormm33} \\
NGC4395  &  ---    &  Sm         &         & --7.27 & $<8.\times10^4$  & \cite{ho98} \\
\noalign{\smallskip}
\hline
\hline
\end{tabular}
}
\end{minipage}

At the risk of greater uncertainty, more galaxies can be included by
combining ground--based stellar kinematics with {\it HST} central light
profiles.  A simple modeling recipe based on two--integral axisymmetric
models has been used on such data for a further 25 E and S0 galaxies
\cite{demogpaper}.  This procedure assumes that the phase--space
density is only a function of the energy and one component of angular
momentum.  MDO masses from this technique can be checked against
galaxies with HST spectroscopy.  The results for 5 low--mass galaxies
with steep inner light profiles show good agreement, but the
2--integral method may overestimate masses by a factor of a few for
massive ellipticals like M87, which rotate slowly and have shallow
central light profiles.  A Bayesian analysis of this sample indicates
that MDOs are in fact very common features of normal, bright galaxy
centers \cite{demogpaper}.  

All known MDOs with measured
masses so far are in galaxies with identifiable spheroidal components,
suggesting that black hole formation is {\it exclusively} linked to
spheroid formation.  However, several well--imaged low z quasars do not
appear to be associated with spheroids \cite{bkss}.

It is clearly important to survey more late--type spirals {\it without}
bulges.  So far, we have little {\it dynamical} evidence on the centers of
such galaxies, but AGN activity might perhaps be used as a proxy 
for a black hole.  Seyferts are generally not found in late--type
spirals \cite{hfs97}, but a single dwarf Seyfert nucleus (out of
hundreds surveyed) has been discovered in NGC 4395, a nearby bulgeless
Sd IV galaxy The bolometric Seyfert luminosity of this nucleus is $1.4
\times 10^{40} \ {\rm ergs/sec}$ \cite{moran98}, and the Eddington BH
mass is only $110 \Msun$, small enough to have been produced by
stellar evolution.  

There is no evidence for MDOs in low--surface brightness galaxies,
although existing studies do not set compelling limits \cite{deblok}.

Our view of these observational results, largely developed over the
past decade, is as follows: (i) MBHs are a normal
feature of the central regions of bright galaxies, particularly those
with spheroids; (ii) their masses scale in rough proportion to
host--galaxy spheroid mass; and (iii) the total mass density in black
holes is broadly consistent with the mass--equivalent energy density in
the quasar light background.  We therefore believe that the black hole
fossils of the quasar era have been found.

\section{Co--evolution of Galaxies and Black Holes}

\subsection{The Era of Quasars}

The improving statistics on local MBHs can be compared to their
properties and distribution during the quasar era at $z \sim 3$.  Were
today's MBHs already fully formed by that time, or was the average MBH
smaller then, having grown by later accretion or mergers to form the 
present population?  The evidence is not conclusive but seems to
favor some growth. 

The epoch of maximal quasar activity in the Universe peaked at the
same time, or slightly before, the epoch of maximal star formation,
and MBHs must have formed {\it before} this time to be available to
power quasars.  Figure 2 illustrates this point by plotting the
history of the rate of observed star formation in the Universe
\cite{madau,connolly,pettini1} together with the density of luminous
quasars (those with $L > 3 \times 10^{46} \, {\rm ergs/sec}$,
\cite{ssg}).  The rise in starbirth is tracked closely by the rise of
luminous quasars.  However the bright quasars reach their peak at $z
\gtsim 2$ ($t \le 1.6 \times 10^9 \, \yr$), and then proceed to die
off nearly $10^9 \, \yr$ before the peak in star formation, which
occurs at $z \sim 1.2$ ($t = 2.6 \times 10^9 \, \yr$)
\cite{boyleterle}.  The application of extinction corrections to the
cosmic star formation rate before $z \sim 2$ is a controversial
subject that may evolve rapidly as better IR data become available.

\begin{figure}
\vskip 0in
\hskip 0.3in\vbox{\psfig{figure=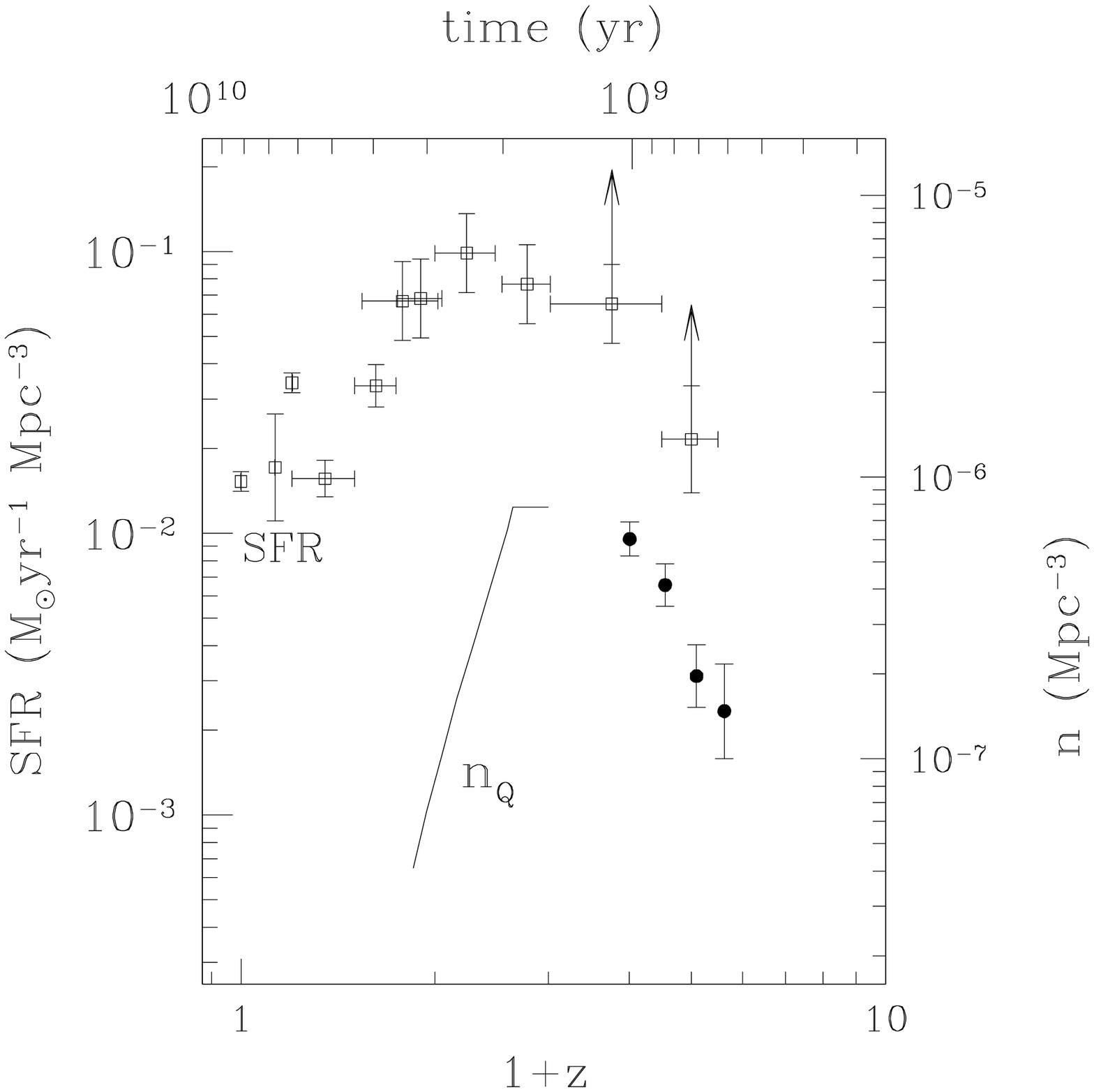,height=5.in,angle=-0.}}
\vskip -0.in
Figure 2. --- 

A comparison of the density of very bright quasars in the Universe,
with the density of star formation, as a function of redshift and
cosmic time.  The solid line and the filled circles are the comoving
number density of quasars ($n_Q$ see [12]) in units of $\mpc^{-3}$,
while the open circles represent estimates of the comoving star
formation {\it rate} (SFR). The SFR can be read off the left axis, and
$n_Q$ should be read off the right axis.  The arrows on the two
highest redshift SFR points indicate an estimate of plausible
correction for extinction.  The peak of quasar activity in the
Universe appears to predate most of the star formation.  The times 
are derived from the redshift assuming $\Omega = 1, \ho = 80
\kms/\mpc$.  

\end{figure}

This chronology favors models in which the black hole forms before, or
in close association with, the densest parts of galaxies (for example
\cite{haehnelt93}), as opposed to models in which the galaxy forms
first and later spawns the quasar.  For this reason we associate the
birth of quasars with {\it spheroid} formation, a process also
identified with dense regions that collapse early \cite{blumenthal84}.
The collapse of these regions would predate the average rise of star
formation illustrated in Figure 2.


The decline of bright quasars at $z<2$ seems likely to 
reflect a loss of fuel.
Galaxy mergers, an effective gas transport process, are  
less frequent as time passes and involve a lower mean density
and slower dynamical clock.  There is also less gas in galaxies overall
(especially spheroids). Radial transport instead may be choked by the
onset of chaos, which would make the spheroid axisymmetric
\cite{merchaos} or through the creation of an inner Lindblad resonance
\cite{sellwood}. Any or all of these effects could play a role in the
reduction of fuel for the black hole.

\def\eps1{\epsilon_{0.1}}

A model in which an MBH forms in every collapsing spheroid invites
comparison between the densities of luminous quasars and galaxies at $z
\sim 2$, and of fossil MBHs today.  
To make this comparison, we must identify a quasar of specified 
luminosity with its fossil MBH mass. We do this under the assumption 
that the brightest quasars are Eddington--limited, that is, that 
their luminosity is so great that radiation pressure on nearby
electrons balances the gravitational force on associated protons.  
In this situation, the ``Eddington luminosity'' is 
\begin{equation}
L_E = 1.3 \times 10^{46} \, M_8 \, {\rm ergs/sec}, 
\end{equation}
where $M_8$ is the mass in units of $10^8 \Msun$.

The comoving density of luminous quasars with $L \ge
6 \times 10^{45} \, {\rm ergs/sec}$ reached its peak value of $10^{-6}
\ \mpc^{-3}$ at $z \sim 3$ \cite{ssg}.  We assume for the moment that
the MBHs underlying these bright quasars {\it have not grown}
substantially since then.  To estimate the corresponding mass range,
we first correct the absolute magnitude limit in the QSO surveys ($M_B
< - 26.0, $ \cite{ssg,boyle}) downward to our $\ho$ and upward by a
bolometric correction factor of 10 \cite{sanders} to get a lower--limit
luminosity of $ 6 \times 10^{46} \, {\rm ergs/sec}$.  This luminosity
implies a MBH mass of $M \sim 4 \times 10^8 \, \Msun$,
corresponding to a bulge luminosity of current hosts (from Figure 2)
of $\sim 10^{10} \, \Lsun$.  Correcting for the  total--to--bulge
luminosity ratio of $3$ \cite{simien}, and using the luminosity function of
bright galaxies \cite{loveday}, we identify the bright
quasars of $z \gtsim 2$ with the half of modern galaxies with $M_B <
-20.7$ that have bulges \cite{bingelli}, we find a 
 comoving density of such spheroids today
of  $\rho_G = 10^{-3} \, \mpc^{-3}$.


Thus, luminous QSO MBHs at $z \sim 3$ are only about $10^{-3}$ as 
numerous as their galaxy host descendents today.  One
way to resolve this discrepancy is to assume that QSOs have very short
duty cycles in their bright phase, of order $\eta \sim 10^{-3}$.
Since the quasar epoch runs about $t_Q \sim 10^9 \, \yr$ from $z
\sim 1.5$ to $z \sim 3.5$ (the FWHM of the quasar plot in Figure 2),
the lifetime in the bright phase would then be only $t_{up} \sim
\eta t_Q = 10^6 \, \yr$.  The fractional mass change
in this phase is only $\delta M/M = t_{up}/t_S \sim 0.02 \, \eps1$,
where the  ``Salpeter  time'' $t_S$ (for an accreting  
black hole to e--fold in mass) is 
\begin{equation}
t_S = M/\dot M = 4 \times 10^7 \, \eps1 \, \yr ,    
\end{equation}
where we have parameterized the radiative efficiency in terms of 
$\eps1 = \epsilon/0.1$ because 
popular geometrically--thin optically--thick accretion disk 
models  rarely exceed  efficiencies of  $\eps1 = 1$. 

This leads to the disturbing conclusion that QSOs accrete only a tiny
fraction of their mass while in their bright phase. This result, which
depends critically on comparing the upper end of the AGN luminosity
function with the upper end of the present--day MBH mass spectrum, is
in sharp contrast with the near equality of the integral quantities
(see Figure 1 \& eqn 3):
the total AGN background is roughly consistent with the total mass
density in present--day MBHs, underpredicting the latter by a modest
factor of 5 (consistent with significant accretion occurring as an
advection dominated flow \cite{narayan98}). Since this integral
constraint is dominated by bright objects, this conflict probably 
reflects a misidentification of the current fossil masses of the 
bright quasars.  

A plausible explanation is that MBHs may {\it not} have stayed
constant in mass from the QSO era until now, but rather grew in mass
by an average factor $F$.  This growth might occur because
hierarchical clustering merged these MBH's along with their
protogalaxies.  These mergers need not emit light.  In that case,
luminous quasars should be identified with MBHs today that are larger
by the factor $F$, and their spheroids would be brighter by nearly the
same factor.  The exponential cutoff in the bright end of the
luminosity function makes such spheroids much rarer and closer in
abundance to the space density of QSOs.

Specifically, for $F = 5$ the limiting MBH mass today becomes $2 \times
10^9 \, \Msun$, the limiting spheroid luminosity is $4 \times 10^{10}
\, \Lsun$, and the new benchmark galaxy in the luminosity function has
$M_B = -22.2$, which is 300 times rarer.  The difference in space
density between quasars and spheroids is reduced to a factor of 3, the
duty cycle comes up to $1/3$, and the lifetime of the bright phase is
$\sim 3 \times 10^8 \, \yr$.  In this model bright quasars now spend a
few Salpeter lifetimes in the bright phase, which is more plausible.

Is growth by a factor $F \sim 5$ reasonable? Data on the recently
discovered Lyman--Break Galaxies (LBGs) indicate slightly more growth.
LBGs have been suggested by several authors as the early--formation
phase of spheroidal components \cite{steidel,lowenthal,trager}.  Small
radii ($1 - 2 \kpc$ \cite {giavalisco,lowenthal}) and small velocity
dispersions (measured for only a handful of the brightest objects
\cite{pettini2}) indicate modest masses of order a few $\times 10^9 \,
\Msun$.  If these merge to form typical spheroids of today with $L
\sim 10^{10} \, \Lsun$ and masses $M \sim 5 \times 10^{10} \, \Msun$,
growth by more like a factor of 10 would be required.  Models of
hierarchically clustering protogalaxies \cite{somerville} also
suggest growth by $F \sim 10$.

A second item that favors more growth is the lack of detection of AGNs
in LBGs.  If MBHs are forming everywhere in protogalaxies together with
stars at the universal ratio $M_{\bullet}/M_{star} = 0.005$ (Figure
1 ), then the average bright LBG should contain an MBH of about $10^7 \,
\Msun$, with Eddington luminosity $L_E \sim 10^{45} \, {\rm
ergs/sec}$.  The apparent magnitude of such an object at $z = 3$
(including absorption by the factor 3--10 that affects LBGs
\cite{meurer}) is about $m_B = 26-27$ , arguably faint
enough to have escaped spectroscopic detection so far.

\subsection{The Era of Galaxies}

If both MBHs and mergers are common among spheroids, two galaxies with
pre--existing MBHs will frequently merge.  The MBHs will spiral towards
the center of the merger remnant, heating and perhaps ejecting stars
from the center \cite{begelrees}.  It is observed \cite{lauerprof}
that low--luminosity ellipticals and spiral bulges possess steeply
rising central star profiles that approximate power laws, while
high--luminosity ellipticals have central profiles that turn over to
form much less dense centers (termed ``cores'').  The dense power--law
centers of the small spheroids, their disky isophotes, and rapid
rotation suggest that gas--rich, dissipative mergers or collapse may
have controlled their mass distributions \cite{nieto91,bender92,
faber4}, perhaps in the presence of one or more MBHs.

The formation of the low--density cores of luminous core galaxies is
more difficult.  Cosmological simulations tend to produce dark matter
distributions with dense centers \cite{navarro}, and dissipation is
likely to sharpen their centers as would adiabatic compression by
accretion of mass by a central black hole. 
\cite{quinhern,faber4}
In the {\it dissipationless } merger of gas--poor galaxies, 
the orbits of the 
associated MBHs would scour out a core in the stellar mass
distribution, with a mass that tracks the black hole mass
\cite{quinhern}. The predicted core masses and radii are a fair match
to those observed, assuming that each spheroid has a BH given by the
standard mass ratio $M_{\bullet}/M_{star} = 0.005$ \cite{faber4}.  The
dissipationless mergers envisioned here are consistent with other
characteristics of core galaxies such as their low rotation and boxy
isophotes.  An additional virtue of MBHs in core galaxies is that they
may protect the cores from infilling by accreting low--mass,
high--density satellites (by tidally shredding them), which may
otherwise give luminous galaxies very bright centers.

\section{Tests of the Picture}

The detection of supermassive black holes and the discovery of
dark matter share a common feature. In both cases there was
skepticism of dynamical mass measurements, and acceptance was preceded
by decades of debate.  Dark matter has since come to be an essential
feature of our understanding of many phenomena, ranging from 
galaxy rotation curves to the formation of structure in the Universe.  
We now seem engaged in a similar transition in the prevailing
view of the centers of galaxies.  

An inevitable source of fuel for dead quasar engines is the debris
from tidally disrupted stars.  MBHs with masses $\mbh \ltsim
10^8M_\odot$ disrupt main--sequence stars rather than swallowing them
whole.  Some of the debris from the star is ejected, but a portion
remains bound to the MBH, forming an accretion disk that undergoes a
``flare'' lasting a few months to a year \cite{goodlee,Rees88}.
Plausible models \cite{Ulmer98} predict a V--band luminosity of about
$10^9L_\odot$.
The event  rate is controlled by how quickly stars can drift into
the ``loss cone'' of low--angular--momentum orbits that come close to
the MBH.  Calculations show that faint, compact
galaxies (e.g., M32) have the highest disruption rates, about
$10^{-4}\yr^{-1}$.  Larger, more diffuse galaxies (e.g., M87) have
much lower rates of about $10^{-6}\yr^{-1}$, and often have
sufficiently massive black holes to consume main sequence stars whole.
    It is possible that a stellar disruption by an MBH has already
been witnessed spectroscopically. The nucleus of the spiral galaxy
NGC~1097  exhibited an ephemeral, broad, double--peaked
H$\alpha$ emission line \cite{F2}, whose profile matched that
expected from an accretion disk \cite{F3}.

Finally, it may be possible to detect the gravitational wave signature
of merging MBHs, and thereby constrain the merger history of galaxies.
In hierarchical models a typical bright galaxy has merged a few times
since the quasar era.  The timescales for decay of binary black holes
in different regimes indicate that, for MBH with $\mbh \gtsim 10^7
\Msun$, the binary holes will merge on a timescale short compared to
the next merger time \cite{xuostriker}.  The merger rate for galaxies
above $0.01 L^*$ may exceed $1/\yr$ in the visible Universe.  For an
equal--mass binary black hole, the final orbit produces a luminosity of
order $L_{grav} \sim c^5/G = 10^{60} {\rm ergs/sec}$ in gravitational
radiation, independent of $\mbh$.  These mergers are the most powerful
events in the Universe, but ironically they may not produce 
electromagnetic radiation.  The energy is emitted over a time $t
\propto G\mbh /c^3$.  The distinctive signature of a supermassive MBH
merger as opposed to two stellar--mass black holes is the lower
frequency and longer duration.  Two $10^7 \Msun$ black holes radiate
much of their energy at a frequency about $10^{-4} \, \hz$.  These events
are too slow for LIGO (the Laser Interferometry 
Gravitational-wave Observatory, but are easy for LISA (the Laser
Interferometric Space Array proposed as a Cornerstone Mission for
ESA).  A key test of the ideas in this paper is the observation of
gravitational radiation from merging black holes at the centers of
merged galaxies since $z \sim 3$.



We thank Sofia Kirhakos for providing images of quasar host galaxies
from reference \cite{bkss}, and Pawan Kumar and John Bahcall for
useful discussions. We thank 
Carl Grillmair for years of fruitful discussions.  Much of this
article depends on research done with the Hubble Space Telescope. We
have enjoyed considerable financial support from NASA. DR acknowledges
the generous support of the J. S. Guggenheim Foundation and the
Ambrose Monell Foundation while at the IAS.

Author Affiliations: 
D. Richstone, Dept of Astronomy, Univ of Michigan {\bf dor@umich.edu};  
E. A. Ajhar, NOAO; 
R. Bender, Ludwig--Maximilians University, Munich;  
G. Bower, NOAO; 
A. Dressler, Observatories of the Carnegie Institution of Washington;  
S. M. Faber, Lick Observatory, University of California;  
A. V. Filippenko, Dept. of Astronomy, U. C. Berkeley;   
K. Gebhardt, Lick Observatory, University of California;  
R. Green, NOAO; 
L. C. Ho, Harvard--Smithsonian Center for Astrophysics;  
J. Kormendy, Institute of Astronomy, Univ of Hawaii;  
 T. Lauer, NOAO; 
J. Magorrian, Canadian Institute of Theoretical Astronomy; 
S. Tremaine, Dept. of Astrophysical Sciences, Princeton U.

\bigskip


\begin{thebibliography}{99}



\def\apj{{\it Astrophys.~J.~}} \def\apjl{{\it
Astrophys.~J. Letters.~}} \def\aj{{\it Astron.~J.~}} \def\aa{{\it
Astron.~Astroph.~}} \def\annrev{{\it Ann.~Rev.~Astron.~\& Astroph.~}}
\def\mn{{\it Mon.~Not.~R.~Astr.~Soc.~}} \def\pasp{{\it Pub.~A~S~P~}}
\def\nature{{\it Nature~}} \def\newa{{\it New Astronomy~}}


\bibitem{begelrees} %
  Begelman, M. C. \& Rees, M. J. Gravity's Fatal Attraction, 
    (W. H. Freeman \& Co., NY 1996).  


\bibitem{kr} Kormendy, J. \& Richstone, D. 
Inward bound --- the search for massive black holes in galactic
nuclei. 
\annrev {\bf 33,} 581 --- 624 (1995). 

\bibitem{kafatos} Kafatos, M. (ed.)  
  ``Supermassive Black Holes'', (Cambridge U Press, Cambridge 1988). 


\bibitem{barr} 
Barr, P. \& Mushotzky, R.~F.
Limits of X-ray variability in active galactic nuclei.
\nature {\bf 320}, 421 --- 423 (1986).  

\bibitem{edelson} 
 Edelson, R. \et  ~ 
Multiwavelength observations of short--timescale variability in NGC~4151. 
IV. Analysis of multiwavelength continuum variability.  
\apj {\bf 470,} 364 --- 377 (1996).  


\bibitem{lbnature} %
  Lynden-Bell, D. 
 Galactic nuclei as collapsed old quasars. 
\nature {\bf 223,} 690 --- 694 (1969). 

\bibitem{blandagn} %
 Blandford, R. D. Physical process in active galactic nuclei. 
in Active Galactic Nuclei 
 (ed. Blandford, R. D., Netzer, H. \& Woltjer, L.)
 (Springer-Verlag, Berlin 1990), pp 161 --274. 

\bibitem{abrambh} %
 Abramowicz, M. A., Bjornsson, G. \& Pringle, J. E. %
 The Theory of Black Hole Accretion Discs
 (Cambridge University Press, Cambridge  1998). 
 
\bibitem{bkss} %
 Bahcall, J. N., Kirhakos, S., Saxe, D. H. \& Schneider, D. P. %
Hubble space telescope images of a sample of 20 nearby luminous
 quasars.  \apj {\bf 479,} 642 - 658 (1997). 

\bibitem{zeld} %
    Zeldovich, Ya. B. \& Novikov, I. D. 
An estimate of the mass of a superstar.   
 {Doklady Akad. Nauk. SSSR} %
     164, {\bf 158,} 311  ---  315 (1964). 


\bibitem{salp} %
  Salpeter, E. E., 
Accretion of interstellar matter by massive objects. 
\apj {\bf 140,} 796  ---  799 (1964).  


\bibitem{lbscripta} %
 Lynden-Bell, D.
 Gravity Power.   
 Physica Scripta {\bf 17,} 185 --- 191 (1978).  



\bibitem{chokshi} %
Chokshi~A. and Turner, E. L.
Remnants of the quasars. 
\mn {\bf 259,} 421 --- 424 (1992).  



\bibitem{loveday} 
Loveday, J.,   Peterson, B. A., Efstathiou, G. 
and Maddox, S. J.  
The Stromlo--APM redshift survey. I. The luminosity function 
and space density of galaxies.  \apj {\bf 390,} 338 --- 344 (1992). 



\bibitem{sargentyoung} %
 Sargent, W. L. W. \et 
Dynamical evidence for a central mass concentration in the 
galaxy M87. 
\apj {\bf 221,} 731 --- 744 (1978).  


\bibitem{drm31}  
Dressler, A. \& Richstone, D. O. 
Stellar dynamics in the nuclei of M31 and M32: evidence for 
massive black holes. 
\apj {\bf 324,} 701 --- 713 (1988). 

\bibitem{kormm31} 
    Kormendy, J.
Evidence for a supermassive black hole in the nucleus of M31. 
\apj {\bf 325,} 128 --- 141 (1988).  


\bibitem{schmeth} 
Schwarzschild, M. 
A dynamical model for a triaxial stellar system in 
dynamical equilibrium.  
\apj {\bf 232,} 236 --- 247 (1979).  


\bibitem{maxent1} Richstone, D.~O. \& Tremaine, S. %
Maximum entropy models of galaxies.  
 \apj  {\bf 327,}  82 --- 88 (1988). 








\bibitem{harmsm87}
   Harms, R. \et 
HST FOS spectroscopy of M87: evidence for a disk of ionized gas and a
   massive black hole.  
\apjl {\bf 435}, L35 ---L38 (1994). 

\bibitem{fordm87} Ford, H. C. \et 
Narrowband HST images of M87: evidence for a disk of ionized gas and a
massive black hole.  
\apjl  {\bf 435,} L27 --- L30 (1994). 

\bibitem{maccm87} %
The supermassive black hole of M87 and the kinematics of its 
associated gaseous disk. 
Macchetto, F. \et 
\apj {\bf 489}, 579 --- 600 (1997).  


\bibitem{youngsargent} 
 Young, P. J., Westphal, J. A., Kristian, J., Wilson, C. P. %
   \&  Landauer, F. P. 
Evidence for a supermassive object in the nucleus of the 
galaxy M87 from SIT and CCD area photometry.  
\apj {\bf 221,} 721 --- 730 (1978).  


\bibitem{miyoshi}
Miyoshi, M. \et 
Evidence for a black hole from high rotation velocities in a
sub--parsec region of NGC 4258.  
\nature {\bf 373}, 127 --- 129 (1995). 

\bibitem{maoz} %
Maoz, E.
  A stringent constraint on alternatives to a massive black hole at
  the center of NGC~4258.
\apjl {\bf 447}, L91 --- L94 (1995).  %
Updated in Maoz, E.
  Dynamical constraints on alternatives to supermassive black holes
  in galactic nuclei.
\apjl {\bf 491}, L181 --- 184 (1998).


\bibitem{genzel} %
  Genzel, R. \& Eckart, A. 
A massive black hole at the center of the Milky Way.   
C. R. Acad. Sci. Paris, 
{\bf 326,} Serie II b, 69 --- 78 (1998). Also, 
Genzel, R. Eckart, A. Ott, T. \& Eisenhauer, F. 
On the nature of the dark mass in the centre of 
the Milky Way.  
\mn {\bf 291,} 219 --- 234 (1997).    

\bibitem{ghez}
Ghez, A. M., Klein, B. L., Mccabe C., Morris, M., 
\& Becklin, E. E. 
High proper motions in the vicinity of SGR A$^*$. 
in the Proceedings of IAU Symposium 184, 
Kyoto, The Central Region of the Galaxy and Galaxies
(ed. Y. Sofue) Kluwer, Dordrecht (1998).  

\bibitem{goodlee}
Goodman, J. \& Lee, H.-M.
Black holes or dark clusters in M31 and M32?
\apj {\bf 337}, 84 --- 90 (1989).   



\bibitem{netzpeter} %
 Netzer, H. \& Peterson, B.~M. 1997, in Astronomical Time Series, ed. 
   D. Maoz, A. Sternberg \& E.~M. Leibowitz (Dordrecht: Kluwer), 85 --- 108


\bibitem{ho98}
Ho, L.~C. ``Supermassive Black Holes in Galactic Nuclei: Observational Evidence and 
Some Astrophysical Consequences.''
In {\it Observational Evidence for Black Holes in the 
Universe}, ed. S.~K. Chakrabarti (Dordrecht: Kluwer), 157 -- 187 
(1998). 


\bibitem{kaspi} %
 Kaspi, S. \et
Multiwavelength observations of short -- timescale variability in NGC~4151. 
II. Optical observations.  
\apj {\bf 470}, 336 --- 348 (1996).




\bibitem{tanaka} %
 Tanaka, Y. \et
Gravitationally redshifted emission implying an accretion
disk and massive black hole in the active galaxy MCG-6-30-15.
\nature {\bf 375}, 659 --- 661 (1995).

\bibitem{nandra} %
 Nandra, K., George, I.~M., Mushotzky, R.~F., Turner, T.~J. \& Yaqoob, T.
ASCA observations of Seyfert 1 galaxies. II. Relativistic Iron K$\alpha$
emission.
\apj {\bf 477}, 602 --- 622 (1997).

\bibitem{reynoldsbeg} %
  Reynolds, C.~S. \& Begelman, M.~C. 
Iron fluorescence from within the innermost stable orbit of black hole
accretion disks.
\apj {\bf 488}, 109 --- 118 (1997).

\bibitem{bromley}%
 Bromley, B. C., Miller, W. A. \& Pariev, V. I. 
The inner edge of the accretion disk around a supermassive black hole.
\nature {\bf 391}, 54 --- 56 (1998).



\bibitem{k93madrid} %
    Kormendy, J. in The Nearest Active Galaxies, ed.~J.~Beckman,
      L.~Colina \& H.~Netzer (Madrid: Consejo Superior de Investigaciones
      Cient\'\i ficas) 197  --- 218 (1993). 


\bibitem{demogpaper}%
   Magorrian J. \et 
The demography of massive dark objects in galaxy centers.
\aj  {\bf 115}, in press (1998).


\bibitem{hfs97}
  Ho, L.C., Filippenko, A.V. \& Sargent, W.L.W.
A search for ``dwarf'' Seyfert nuclei. V. Demographics of nuclear
                            activity in nearby galaxies.
\apj {\bf 487}, 568 --- 578 (1997).

\bibitem{moran98} 
  Moran, E. C. \et
X-Rays from NGC 4395, the least luminous Seyfert 1 nucleus.
Submitted to \apj


\bibitem{deblok}  %
de Blok, W.J.G., McGaugh, S.S. \& van der Hulst, J.M.
HI observations of low surface brightness galaxies: probing low-density
galaxies.
\mn {\bf 283}, 18 --- 54 (1996).

\bibitem{madau}  %
Madau, P., Ferguson, H. C., Dickinson, M. E., Giavalisco, M., 
  Steidel, C. C. \& Fruchter, A.  
High-redshift galaxies in the Hubble Deep Field: Colour selection and star
formation history to $z\sim4$.
\mn {\bf 283}, 1388 --- 1404 (1996).


\bibitem{connolly} %
  Connolly, A. J., Szalay, A. S. Dickinson, M., Subbarao, M. U. \& 
  Brunner, R. J.
 The evolution of the global star formation history as measured from
                        the Hubble Deep Field.
\apjl {\bf 486}, L11 --- L14 (1997).

\bibitem{pettini1}  %
Pettini, M., Steidel, C.C., Adelberger, K.L., Kellogg, M.,
Dickinson, M. \&  Giavalisco, M. 
in ``Cosmic Origins: Evolution of Galaxies, Stars, Planets and Life'' 
ed. J.M. Shull, C.E. Woodward, and H.A. Thronson, 
(ASP Conference Series), in press (1998).


\bibitem{ssg} 
 Schmidt, M., Schneider, D. P. \& Gunn, J. E. 
Spectroscopic CCD surveys for quasars at 
large redshift.  IV. Evolution of the luminosity 
function from quasars detected by their Lyman-alpha 
emission. 
  \aj {\bf 110,} 68 --- 77 (1995). 


\bibitem{boyleterle}
Boyle, B. J., Terlevich, R. J. 
The cosmological evolution of the QSO luminosity density 
and of the star formation rate. 
 \mn {\bf 293}, L49 --- L51 (1998). 

\bibitem{haehnelt93} %
Haehnelt, M. G.  and Rees, M. J.
The formation of nuclei in newly formed galaxies and the evolution
of the quasar population.
\mn {\bf 263}, 168 --- 178 (1993). %
See also Haehnelt, M. G., Natarajan, P. \& Rees, M. J. %
High-redshift galaxies, their active nuclei and central black holes.
Submitted to \mn (1997).

\bibitem{blumenthal84}
 Blumenthal, G.R., Faber, S.M., Primack, J.R., Rees, M.J.
     Formation of galaxies and large-scale structure with cold dark matter.
     \nature  {\bf 311}, 517 --- 525 (1984). 



\bibitem{merchaos} %
Merritt, D. \& Quinlan, J. D. 
Dynamical evolution of elliptcal galaxies with central singularities. 
\apj {\bf 498}, 625 -- 639 (1998)


\bibitem{sellwood} Sellwood, J. A. \& Moore, E. M. 
On the formation of disk galaxies and massive central objects. 
\apj ~ submitted (1998).   


\bibitem{boyle} 
Boyle, B.J., Jones, L.R., Shanks, T.,
Marano, B., Zitelli, V.  \& Zamorani, G.
``QSO evolution and clustering at $z<2.9$'' in 
The Space Distribution of Quasars (ed D. Crampton)
pp. 191 --- 198 (San Francisco: Astron.  Soc. Pac, 1991).
See also 
Boyle, B. J. Shanks, T. \& Peterson, B. A. 
The Evolution of Optically Selected QSOs -- II.   
 \mn {\bf 235}, 935 --- 948 (1988).



\bibitem{sanders}
  Sanders, D. B., Phinney, E. S., Neugebauer, G., Soifer, B. T.
  \& Matthews, K. 
Continuum energy distributions of quasars: shapes and origins. 
 \apj {\bf 347}, 29 --- 51 (1989).


\bibitem{simien} 
Simien, F. \& de Vaucouleurs, G.  
Systematics of bulge-to-disk ratios.  
\apj {\bf 302}, 564 --- 578 (1986).  


\bibitem{bingelli}  
Binggeli, B., Sandage, A. \& Tammann, G. A.
The luminosity functions of galaxies. 
\annrev {\bf 26}, 509 --- 560 (1988). 


\bibitem{narayan98} 
 Narayan, R., Mahadevan, R., and Quataert, E. 
   in ``The Theory of Black Hole Accretion Discs'', eds Abramowicz, 
   Bjornsen \& Pringle (Cambridge U Press) 1988, and also 
   astro-ph/9803141.  


\bibitem{steidel}
Steidel, C.~C., Giavalisco, M., Pettini, M., Dickinson, M. \&
  Adelberger, K.~L.
Spectrosopic confirmation of a population of normal star-forming
 galaxies at redshifts $z>3$. 
\apjl {\bf 462}, L17 --- L21 (1996).


\bibitem{lowenthal}
    Lowenthal, J.D. \et
     Keck spectroscopy of redshift $z\sim3$ galaxies in the Hubble deep
     field.
     \apj {\bf 481}, 673 --- 688 (1997).


\bibitem{trager}
   Trager, S.C., Faber, S.M., Dressler, A. \& Oemler, A. 
Galaxies at $z\approx4$ and the formation of Population~II.
   \apj {\bf 485}, 92 --- 99 (1997).


\bibitem{giavalisco}
 Giavalisco, M., Steidel, C.C. \& Machetto, F.D.
Hubble Space Telescope imaging of star-forming galaxies at
redshifts $z>3$.
\apj {\bf 470}, 189 --- 194 (1996).

\bibitem{pettini2}
 Pettini, M. \et . 
Infrared observations of nebular emission lines from galaxies at z=3. 
\apj ~ in the press (1998).   



\bibitem{somerville}
Somerville, R., Primack, J. \& Faber, S.M. 
The nature of high redshift galaxies. 
\apj ~ submitted (1998).  


\bibitem{meurer}
Meurer, G., Heckman, T., Lehnert, M., Leitherer, C. \& Lowenthal, J.
The panchromatic starburst intensity limit at low and high
redshift.
\aj {\bf 114}, 54 --- 68 (1997).

\bibitem{lauerprof}
 Lauer, T.R. \et
The centers of early-type galaxies with HST. I. An
observational survey.
\aj {\bf 110}, 2622 --- 2654  (1995).


\bibitem{nieto91} %
Nieto, J.-L., Bender, R. \& Surma, P.
Central brightness profiles
and isophotal shapes in elliptical galaxies.
\aa {\bf 244}, L37-L40 (1991).

\bibitem{bender92}
Bender, R., Burstein., D. \& Faber, S.M.
Dynamically hot galaxies. I -- Structural properties.
\apj {\bf 399}, 462 --- 477 (1992).


\bibitem{faber4} 
Faber, S.M. \et
The centers of early-type galaxies with HST. IV. Central parameter relations.
\aj {\bf 114}, 1771 --- 1796 (1997).


\bibitem{navarro}
 Navarro, J.F., Frenk, C. \& White, S.D.M.
The structure of Cold Dark Matter halos.
\apj {\bf 462}, 563 --- 575 (1996).



\bibitem{quinhern} 
 Quinlan, G.D.
 The dynamical evolution of massive black hole binaries I. Hardening in a
 fixed stellar background.
\newa {\bf 1}, 35 --- 56 (1996).

Quinlan, G.D. \& Hernquist, L.
The dynamical evolution of massive black hole binaries II. Self-consistent
$N$-body integrations.
\newa {\bf 2}, 533 --- 554 (1997).

\bibitem{Rees88}%
    Rees M.J., 
Tidal disruption of stars by black holes of $10^6 - 10^8$
solar masses in nearby galaxies.
\nature  {\bf 333}, 523 --- 528 (1988). 



\bibitem{Ulmer98}%
      Ulmer, A.
Flares from the tidal disruption of stars by massive black holes.
\apj, in press (1998).





\bibitem{F2} Storchi-Bergmann, T. \et
The variability of the double-peaked Balmer lines in the active
nucleus of NGC~1097.
\apj {\bf 443}, 617 --- 624 (1995).

\bibitem{F3} Chen, K. \& Halpern, J. P.
Structure of line-emitting accretion disks in active galactic nuclei --
Arp 102B.
\apj {\bf 344}, 115 --- 124 (1989).







\bibitem{xuostriker} 
 Xu, G. \& Ostriker, J. P. 
 Dynamics of massive black holes as a possible candidate of Galactic dark
 matter.
\apj {\bf 437}, 184 --- 193 (1994). 





\bibitem{tonry84} %
  Tonry, J.
Evidence for a central mass concentration in M32.
\apjl {\bf 283}, L27--L30 (1984).


\bibitem{benderm32}
Bender, R., Kormendy, J. \&  Dehnen, W. 
Improved evidence for 
a $3 \times 10^6 \Msun$ Black Hole in M32 
\apjl {464,} L123-L126 (1996).

\bibitem{vdmm32} 
van der Marel, R.~P., Cretton, N., de Zeeuw, P.~T. \&  Rix, H.-W.
Improved evidence for a black hole in M32 from HST/FOS spectra.  
II. Axisymmetric dynamcial models.  
\apj  {\bf 493,} 613-631 (1998).  

\bibitem{krn3115} %
    Kormendy, J. \& Richstone, D. 
Evidence for a supermassive black hole in NGC~3115. 
\apj {\bf 393}, 559 --- 578 (1992).  

\bibitem{knuk3115} %
    Kormendy, J. \et
Hubble Space Telescope spectroscopic evidence for a
$2\times10^9\,M_\odot$ black hole in NGC~3115.
\apjl {\bf 459}, L57 --- L60 (1996).

\bibitem{korm3377} %
   Kormendy, J., Bender, R., Evans, A. \& Richstone, D. 
 The mass distribution in the elliptical galaxy NGC 3377: 
evidence for a $ 2 \times 10^8 \ \Msun $ Black Hole.  
   \aj ~ {\bf 115}, 1823 --- 1839 (1998).  


\bibitem{geb3379} 
  Gebhardt, K. \et
Axisymmetric, three-integral models of galaxies: A massive black hole
in NGC~3379.
\apj submitted (1988).

\bibitem{vdbosch97} 
 Cretton, N. \& van den Bosch, F.~C.
 Evidence for a massive black hole in the S0 galaxy NGC~4342.
 Submitted to \apj  (1998).  astro-ph/9805324. 


\bibitem{knuk4594}
  Kormendy, J. \et
Hubble Space Telescope spectroscopic evidence for a $1\times10^9\,M_\odot$
black hole in NGC~4594.
\apjl {\bf 473}, L91 --- L94 (1996).


\bibitem{kormn4594}
   Kormendy, J.
Evidence for a central dark mass in NGC~4594 (the Sombrero galaxy).
\apj {\bf 335}, 40 --- 56 (1988).   

\bibitem{bowerm84}
Bower, G. A. \et 
Kinematics of the nuclear ionized gas in the radio galaxy M84
(NGC~4374).
\apjl {\bf 492}, L111 --- L114 (1998). 


\bibitem{ferrn4261}
  Ferrarese, L., Ford, H.~C. \& Jaffe, W.
Evidence for a massive black hole in the active galaxy NGC~4261
from Hubble Space Telescope images and spectra.
\apj {\bf 470}, 444 --- 459 (1996).

\bibitem{vdmn7052} 
  van der Marel, R. P. \& van den Bosch, F. C.  
Evidence for a $3 \times 10^8$ solar mass black hole in NGC 7052 from
  HST observations of the nuclear gas disk.  
  \aj ~ submitted (1998).  Astro-ph/9804194


\bibitem{grnhill1068} 
  Greenhill, L.~J., Gwinn, C.~R., Antonucci, R. \& Barvainis, R.
VLBI imaging of water maser emission from the nuclear torus of NGC~1068.
\apjl {\bf 472}, L21 --- L24 (1996). 

\bibitem{grnhill4945} 
  Greenhill, L.~J., Moran, J.~M. \& Herrnstein, J.~R.
The distribution of H$_2$O maser emission in the Nucleus of NGC~4945.
\apjl {\bf 481}, L23 --- L26 (1997).  


\bibitem{jones}
  Jones, D. \et
Visible and far-ultraviolet WFPC2 imaging of the nucleus of
the galaxy NGC~205.
\apj {\bf 466}, 742 --- 749 (1996).  

\bibitem{kormm33} 
  Kormendy, J. \& McClure, R.~D.
The nucleus of M33.
\aj {\bf 105}, 1793 --- 1812 (1993).  




\end{thebibliography}
\end{document}